\pdfoutput=1
\documentclass[12pt,a4paper]{article}

\usepackage{ifthen}
\newboolean{pdflatex}
\setboolean{pdflatex}{true}

\newboolean{articletitles}
\setboolean{articletitles}{true}

\newboolean{uprightparticles}
\setboolean{uprightparticles}{false}

\def\paperauthors{LHCb collaboration}
\def\paperasciititle{ Momentum scale calibration of the LHCb spectrometer}
\def\papertitle{Momentum scale calibration of the LHCb spectrometer}
\def\paperkeywords{{High Energy Physics}, {LHCb}}
\def\papercopyright{\the\year\ CERN for the benefit of the LHCb collaboration}
\def\paperlicence{CC BY 4.0 licence}
\def\paperlicenceurl{https://creativecommons.org/licenses/by/4.0/}

\usepackage[top=1in, bottom=1.25in, left=1in, right=1in]{geometry}

\usepackage{microtype}
\usepackage{lineno}
\usepackage{xspace}
\usepackage{caption}

\usepackage{graphicx}
\usepackage{color}
\usepackage{colortbl}

\usepackage{amsmath}
\usepackage{amssymb}
\usepackage{amsfonts}
\usepackage{upgreek}

\newcommand*\patchAmsMathEnvironmentForLineno[1]{
\expandafter\let\csname old#1\expandafter\endcsname\csname #1\endcsname
\expandafter\let\csname oldend#1\expandafter\endcsname\csname
end#1\endcsname
 \renewenvironment{#1}
   {\linenomath\csname old#1\endcsname}
   {\csname oldend#1\endcsname\endlinenomath}
}
\newcommand*\patchBothAmsMathEnvironmentsForLineno[1]{
  \patchAmsMathEnvironmentForLineno{#1}
  \patchAmsMathEnvironmentForLineno{#1*}
}
\AtBeginDocument{
\patchBothAmsMathEnvironmentsForLineno{equation}
\patchBothAmsMathEnvironmentsForLineno{align}
\patchBothAmsMathEnvironmentsForLineno{flalign}
\patchBothAmsMathEnvironmentsForLineno{alignat}
\patchBothAmsMathEnvironmentsForLineno{gather}
\patchBothAmsMathEnvironmentsForLineno{multline}
\patchBothAmsMathEnvironmentsForLineno{eqnarray}
}

\usepackage{hyperxmp}

\usepackage[pdftex,
            pdfauthor={\paperauthors},
            pdftitle={\paperasciititle},
            pdfkeywords={\paperkeywords},
            pdfcopyright={Copyright (C) \papercopyright},
            pdflicenseurl={\paperlicenceurl}]{hyperref}

\usepackage[colorinlistoftodos,textsize=scriptsize]{todonotes}

\usepackage[bottom,flushmargin,hang,multiple]{footmisc}

\usepackage[all]{hypcap}

\usepackage{xspace}
\usepackage{upgreek}

\def\lhcb   {\mbox{LHCb}\xspace}

\def\MagUp {\mbox{\em Mag\kern -0.05em Up}\xspace}
\def\MagDown {\mbox{\em MagDown}\xspace}

\ifthenelse{\boolean{uprightparticles}}
{

 \def\Pmu         {\ensuremath{\upmu}\xspace}

 \def\Ppi         {\ensuremath{\uppi}\xspace}

 \def\Ppsi        {\ensuremath{\uppsi}\xspace}

 \def\PDelta      {\ensuremath{\Delta}\xspace}
 \def\PXi         {\ensuremath{\Xi}\xspace}
 \def\PLambda     {\ensuremath{\Lambda}\xspace}
 \def\PSigma      {\ensuremath{\Sigma}\xspace}
 \def\POmega      {\ensuremath{\Omega}\xspace}
 \def\PUpsilon    {\ensuremath{\Upsilon}\xspace}
 \let\oldPi\Pi
 \def\PPi         {\ensuremath{\oldPi}\xspace}

 \def\PB      {\ensuremath{\mathrm{B}}\xspace}
 
 \def\PD      {\ensuremath{\mathrm{D}}\xspace}

 \def\PJ      {\ensuremath{\mathrm{J}}\xspace}
 \def\PK      {\ensuremath{\mathrm{K}}\xspace}

 \def\PZ      {\ensuremath{\mathrm{Z}}\xspace}
 
 \def\Pb      {\ensuremath{\mathrm{b}}\xspace}
 \def\Pc      {\ensuremath{\mathrm{c}}\xspace}

 \def\Pi      {\ensuremath{\mathrm{i}}\xspace}

 \def\Ps      {\ensuremath{\mathrm{s}}\xspace}

 \def\thebaroffset{0.0em}
}
{

 \def\Pmu         {\ensuremath{\mu}\xspace}

 \def\Ppi         {\ensuremath{\pi}\xspace}

 \def\Ppsi        {\ensuremath{\psi}\xspace}
 
 \mathchardef\PDelta="7101
 \mathchardef\PXi="7104
 \mathchardef\PLambda="7103
 \mathchardef\PSigma="7106
 \mathchardef\POmega="710A
 \mathchardef\PUpsilon="7107
 \mathchardef\PPi="7105
 
 \def\PB      {\ensuremath{B}\xspace}
 
 \def\PD      {\ensuremath{D}\xspace}

 \def\PJ      {\ensuremath{J}\xspace}
 \def\PK      {\ensuremath{K}\xspace}

 \def\PZ      {\ensuremath{Z}\xspace}
 
 \def\Pb      {\ensuremath{b}\xspace}
 \def\Pc      {\ensuremath{c}\xspace}

 \def\Pi      {\ensuremath{i}\xspace}

 \def\Ps      {\ensuremath{s}\xspace}

 \def\thebaroffset{0.18em}
}
\newcommand{\offsetoverline}[2][\thebaroffset]{\kern #1\overline{\kern -#1 #2}}

\makeatletter
\ifcase \@ptsize \relax
  \newcommand{\miniscule}{\@setfontsize\miniscule{4}{5}}
\or
  \newcommand{\miniscule}{\@setfontsize\miniscule{5}{6}}
\or
  \newcommand{\miniscule}{\@setfontsize\miniscule{5}{6}}
\fi
\makeatother

\DeclareRobustCommand{\optbar}[1]{\shortstack{{\miniscule (\rule[.5ex]{1.25em}{.18mm})}
  \\ [-.7ex] $#1$}}

\def\mup        {{\ensuremath{\Pmu^+}}\xspace}
\def\mun        {{\ensuremath{\Pmu^-}}\xspace}

\def\Z      {{\ensuremath{\PZ}}\xspace}

\def\squark    {{\ensuremath{\Ps}}\xspace}

\def\cquark    {{\ensuremath{\Pc}}\xspace}

\def\bquark    {{\ensuremath{\Pb}}\xspace}

\def\pion   {{\ensuremath{\Ppi}}\xspace}

\def\pip    {{\ensuremath{\pion^+}}\xspace}

\def\kaon    {{\ensuremath{\PK}}\xspace}

\def\KorKbar {\kern \thebaroffset\optbar{\kern -\thebaroffset \PK}{}\xspace}

\def\Kp      {{\ensuremath{\kaon^+}}\xspace}
\def\Km      {{\ensuremath{\kaon^-}}\xspace}

\def\KS      {{\ensuremath{\kaon^0_{\mathrm{S}}}}\xspace}

\def\D       {{\ensuremath{\PD}}\xspace}

\def\DorDbar {\kern \thebaroffset\optbar{\kern -\thebaroffset \PD}\xspace}
\def\Dz      {{\ensuremath{\D^0}}\xspace}

\def\Dp      {{\ensuremath{\D^+}}\xspace}
\def\Dm      {{\ensuremath{\D^-}}\xspace}

\def\DpDm    {\ensuremath{\Dp {\kern -0.16em \Dm}}\xspace}

\def\B       {{\ensuremath{\PB}}\xspace}

\def\BorBbar {\kern \thebaroffset\optbar{\kern -\thebaroffset \PB}\xspace}

\def\Bd      {{\ensuremath{\B^0}}\xspace}

\def\BdorBdbar {\kern \thebaroffset\optbar{\kern -\thebaroffset \Bd}\xspace}
\def\Bu      {{\ensuremath{\B^+}}\xspace}

\def\Bs      {{\ensuremath{\B^0_\squark}}\xspace}

\def\BsorBsbar {\kern \thebaroffset\optbar{\kern -\thebaroffset \Bs}\xspace}

\def\jpsi     {{\ensuremath{{\PJ\mskip -3mu/\mskip -2mu\Ppsi}}}\xspace}

\def\Y#1S{\ensuremath{\PUpsilon{(#1S)}}\xspace}
\def\OneS  {{\Y1S}\xspace}
\def\TwoS  {{\Y2S}\xspace}
\def\ThreeS{{\Y3S}\xspace}

\def\Lz          {{\ensuremath{\PLambda}}\xspace}

\def\LorLbar     {\kern \thebaroffset\optbar{\kern -\thebaroffset \PLambda}\xspace}

\def\Xires       {{\ensuremath{\PXi}}\xspace}

\def\Omegares    {{\ensuremath{\POmega}}\xspace}

\def\Xiccpp      {{\ensuremath{\Xires^{++}_{\cquark\cquark}}}\xspace}

\def\Lb           {{\ensuremath{\Lz^0_\bquark}}\xspace}

\def\Xibm         {{\ensuremath{\Xires^-_\bquark}}\xspace}

\def\Omegab       {{\ensuremath{\Omegares^-_\bquark}}\xspace}

\def\to                 {\ensuremath{\rightarrow}\xspace}

\def\AT#1     {\ensuremath{A_{\mathrm{T}}^{#1}}\xspace}

\def\C#1      {\ensuremath{\mathcal{C}_{#1}}\xspace}
\def\Cp#1     {\ensuremath{\mathcal{C}_{#1}^{'}}\xspace}
\def\Ceff#1   {\ensuremath{\mathcal{C}_{#1}^{\mathrm{(eff)}}}\xspace}
\def\Cpeff#1  {\ensuremath{\mathcal{C}_{#1}^{'\mathrm{(eff)}}}\xspace}
\def\Ope#1    {\ensuremath{\mathcal{O}_{#1}}\xspace}
\def\Opep#1   {\ensuremath{\mathcal{O}_{#1}^{'}}\xspace}

\newcommand{\aunit}[1]{\ensuremath{\text{\,#1}}}

\newcommand{\tev}{\aunit{Te\kern -0.1em V}\xspace}
\newcommand{\gev}{\aunit{Ge\kern -0.1em V}\xspace}
\newcommand{\mev}{\aunit{Me\kern -0.1em V}\xspace}
\newcommand{\kev}{\aunit{ke\kern -0.1em V}\xspace}
\newcommand{\ev}{\aunit{e\kern -0.1em V}\xspace}

\newcommand{\mevc}{\ensuremath{\aunit{Me\kern -0.1em V\!/}c}\xspace}
\newcommand{\gevc}{\ensuremath{\aunit{Ge\kern -0.1em V\!/}c}\xspace}
\newcommand{\mevcc}{\ensuremath{\aunit{Me\kern -0.1em V\!/}c^2}\xspace}
\newcommand{\gevcc}{\ensuremath{\aunit{Ge\kern -0.1em V\!/}c^2}\xspace}

\def\cm   {\aunit{cm}\xspace}

\def\fb   {\ensuremath{\aunit{fb}}\xspace}
\def\invfb   {\ensuremath{\fb^{-1}}\xspace}

\def\deriv {\ensuremath{\mathrm{d}}}

\def\gsim{{~\raise.15em\hbox{$>$}\kern-.85em
          \lower.35em\hbox{$\sim$}~}\xspace}
\def\lsim{{~\raise.15em\hbox{$<$}\kern-.85em
          \lower.35em\hbox{$\sim$}~}\xspace}

\def\ptot       {\ensuremath{p}\xspace}

\def\evtgen     {\mbox{\textsc{EvtGen}}\xspace}

\def\geant      {\mbox{\textsc{Geant4}}\xspace}

\def\photos     {\mbox{\textsc{Photos}}\xspace}

\def\pythia     {\mbox{\textsc{Pythia}}\xspace}

\def\tell1  {TELL1\xspace}
\def\ukl1   {UKL1\xspace}

\newcommand{\lhcborcid}[1]{\href{https://orcid.org/#1}{\hspace*{0.1em}\raisebox{-0.45ex}{\includegraphics[width=1em]{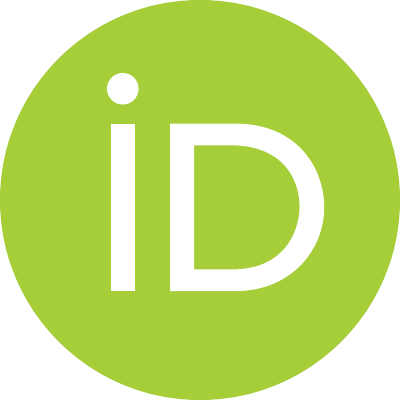}}}}

\usepackage{cite}
\usepackage{mciteplus}

\begin{document}

\renewcommand{\thefootnote}{\fnsymbol{footnote}}
\setcounter{footnote}{1}

\begin{titlepage}
\pagenumbering{roman}

\vspace*{-1.5cm}
\centerline{\large EUROPEAN ORGANIZATION FOR NUCLEAR RESEARCH (CERN)}
\vspace*{1.5cm}
\noindent
\begin{tabular*}{\linewidth}{lc@{\extracolsep{\fill}}r@{\extracolsep{0pt}}}
\vspace*{-1.5cm}\mbox{\!\!\!\includegraphics[width=.14\textwidth]{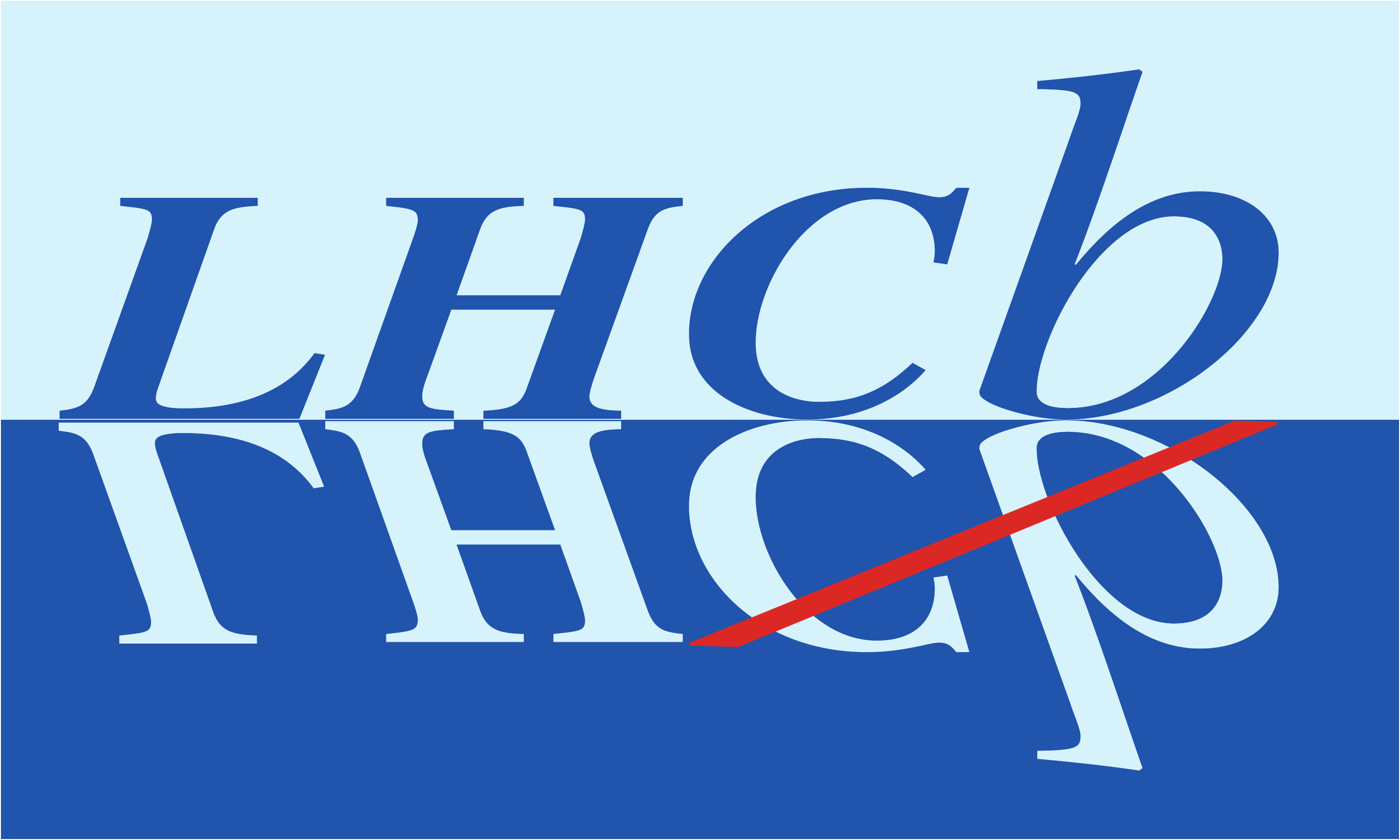}} & &
\\
 & & CERN-EP-2023-275 \\
 & & LHCb-DP-2023-003 \\
 & & $5^{th}$ February, 2024 \\
 & &  \\
\end{tabular*}

\vspace*{4.0cm}

{\normalfont\bfseries\boldmath\huge
\begin{center}
  \papertitle
\end{center}
}

\vspace*{2.0cm}

\begin{center}
\paperauthors\footnote{Authors are listed at the end of this paper.}
\end{center}

\vspace{\fill}

\begin{abstract}
  \noindent
For accurate determination of particle masses accurate knowledge of the momentum scale of the detectors is crucial. The procedure used to calibrate the momentum scale of
the LHCb spectrometer is described and illustrated using the performance obtained with an integrated luminosity of $1.6\invfb$ collected during 2016 in $pp$ running. The procedure uses large samples of $\jpsi \rightarrow \mu^+ \mu^-$  and $\Bu \rightarrow \jpsi K^+$  decays and leads to a relative accuracy of $3 \times 10^{-4}$ on the momentum scale.
\end{abstract}

\vspace*{2.0cm}

\begin{center}
Published in  JINST
\end{center}

\vspace{\fill}

{\footnotesize
\centerline{\copyright~\papercopyright. \href{\paperlicenceurl}{\paperlicence}.}}
\vspace*{2mm}

\end{titlepage}

\newpage
\setcounter{page}{2}
\mbox{~}

\renewcommand{\thefootnote}{\arabic{footnote}}
\setcounter{footnote}{0}

\cleardoublepage

\pagestyle{plain}
\setcounter{page}{1}
\pagenumbering{arabic}

\section{Introduction}
\label{sec:Introduction}
An accurate estimate of the charged-particle momentum scale is a critical
ingredient to achieve optimal performance in
spectrometers such as the LHCb detector \cite{LHCb-DP-2008-001}. If the
momentum scale is not well calibrated, the measured four-vectors of
charged particles will be biased and  the measured
masses of resonances and other quantities such as decay-time will be shifted from their true values. If the bias is large and depends on the particle kinematics
the mass resolution is also degraded, particularly for high mass resonances such as the $\PUpsilon(nS)$.

In this paper the procedure used to calibrate the measurement of the charged-particle momentum scale of the LHCb spectrometer from 2011 to 2018 is described and results are presented using data collected during 2016 as an example. Similar results are obtained for other running periods. The paper is arranged as follows. First, the LHCb spectrometer
and the formalism used are described. The procedure is then
illustrated using a data sample corresponding to an integrated luminosity of $1.6 \invfb$, collected in proton-proton ($pp$) collisions at
a centre-of-mass energy of $13 \tev$ during 2016 data taking. Finally, comparison
is drawn with data collected by LHCb during 2012 running and comments are made on addressing biases related to the correlation between the mass and decay time.

\section{Detector}
\label{sec:Detector}
The \lhcb detector~\cite{LHCb-DP-2008-001,LHCb-DP-2014-002} is a single-arm forward
spectrometer covering the \mbox{pseudorapidity} range $2<\eta <5$,
designed for the study of particles containing \bquark or \cquark
quarks. The detector includes a high-precision tracking system
consisting of a silicon-strip vertex detector (VELO) surrounding the $pp$
interaction region~\cite{LHCb-DP-2014-001}, a large-area silicon-strip detector (TT) located
upstream of a dipole magnet with a bending power of about
$4{\mathrm{\,Tm}}$, and three stations of silicon-strip detectors and straw
drift tubes~\cite{LHCb-DP-2013-003} placed downstream of the
magnet (T-stations). The length scale of the VELO along the beam axis (the
$z$-direction) is known to a relative precision of $2 \times 10^{-4}$. The polarity of the dipole magnet is reversed periodically throughout data-taking.
The configuration with the magnetic field vertically upwards, \MagUp
(downwards, \MagDown), deflects positively (negatively) charged
particles in the horizontal plane towards the centre of the LHC. The
tracking system provides a measurement of the momentum, \ptot, of charged particles with
a relative uncertainty that varies from 0.5\% at low momentum to 1.0\%
at 200\gevc.

Different types of charged hadrons are distinguished using information
from two ring-imaging Cherenkov detectors~\cite{LHCb-DP-2012-003}.
Photons, electrons, and hadrons are identified by a calorimeter system consisting of
scintillating-pad and preshower detectors, an electromagnetic
calorimeter and a hadronic calorimeter. Muons are identified by a
system composed of alternating layers of iron and multiwire
proportional chambers~\cite{LHCb-DP-2012-002}.
The online event selection is performed by a trigger~\cite{LHCb-DP-2012-004},
which consists of a hardware stage, based on information from the calorimeter and muon
systems, followed by a software stage, which applies a full event
reconstruction including a track fit based on a Kalman filter \cite{kalman,FRUHWIRTH1987444}.

In the simulation, $pp$ collisions are generated using
\pythia~6.4~\cite{Sjostrand:2006za} with a specific \lhcb
configuration~\cite{LHCb-PROC-2010-056}.  Decays of hadronic particles
are described by \evtgen~\cite{Lange:2001uf} in which final state
radiation is generated using \photos~\cite{Golonka:2005pn}. The
interaction of generated particles with the detector and its
response are implemented using the \geant
toolkit~\cite{Allison:2006ve, *Agostinelli:2002hh} as described in
Ref.~\cite{LHCb-PROC-2011-006}.

The accuracy of the momentum scale prior to the procedure described
here is determined by the knowledge of the magnetic field map, the
alignment, and the detector
material. The field of the dipole
magnet was mapped to a relative accuracy of $4 \times 10^{-4}$ with Hall
probes prior to first data taking in 2009 \cite{LHCb-DP-2008-001}. Further measurements were
taken during LHC shutdown periods. These measurements were combined with the
results of a TOSCA finite-element simulation to provide the field map used in
the reconstruction.

The geometry of the LHCb tracking system is relatively complicated and
care is needed to achieve the optimal alignment. Detailed metrology and optical surveys were made for all
detector components prior to installation. The stability of the
C-frames on which the T-stations are mounted are monitored using a CCD-based Rasnik system
\cite{LHCb-DP-2013-003}. During 2014 an additional optical monitoring
system (`BCAM') was installed to monitor the position of the inner
part of this detector \cite{Gayde:2667533}. From the information provided by the hardware
system and software alignment studies it is known that the position of most
detector elements is stable during data taking. An exception is the silicon-strip detector located in the inner part of the first T-station. A shift of $\sim 1 \cm$ in the position of this detector along the
beam line is seen when the magnet is powered on.

The framework to determine the global alignment of the tracking system is discussed in Refs.\cite{JAmoraal_2010, Deissenroth_2010}. It is based on the closed form Kalman filter-based method described in Ref.~\cite{Hulsbergen_2009}. An important aspect of the software alignment procedure is that it utilizes mass constraints provided by selected $\Dz\rightarrow K^+
\pi^-$ and $\jpsi \rightarrow \mup \mu^{-}$ candidates to reduce weak modes such as the curvature bias described in Section~\ref{sec:formal} and Ref. \cite{Amoraal:2012qn}. During Run 1 (2010--2012) updates to the
alignment constants were made offline when significant changes to the
running conditions occurred (e.g. changes in magnet polarity). From 2015 onwards (Run 2) the alignment runs in
real-time each fill at the software trigger stage, using the $\Dz$ mass constraint, and the
constants are automatically updated if significant changes are detected \cite{LHCb-PROC-2015-011}. Midway through the 2018 run period a high-granularity alignment of the spectrometer was made using a sample of high-momentum muons produced in $\Z^0 \rightarrow \mu^{+} \mu^{-}$ decays. The resulting alignment was then used as an input to the online procedure.

A correction for energy
loss in the detector material is applied during the Kalman filter step. This correction uses the Bethe formula including the density correction \cite{PDG2022}. As
particle identification information is not available when the track fit is performed, all particles are considered to be pions. This is a valid assumption as particles that traverse
the full spectrometer are highly relativistic ($p > 3 \gevc$) such that the energy loss correction is independent of the particle type. The size of the applied correction is momentum
dependent and comparable to the intrinsic momentum resolution of the
detector. The amount of material in the detector is known from surveys
and assays during installation to a precision of $5- 10  \, \%$
\cite{LHCb-DP-2014-002}. In detailed studies of particles' masses, an additional uncertainty to the momentum scale determination that accounts for the material knowledge is assigned \cite{LHCb-PAPER-2013-011}.

From the above discussion it is clear that to achieve an accurate calibration of the
momentum scale using the decays of precisely measured resonances
are needed. The methodology for this will be detailed in the next section.

\section{Formalism}
\label{sec:formal}
Consider a two-body decay $P \rightarrow d_1d_2$. The invariant mass
of the system, in natural units, is
\begin{equation}
  m_{1 2} = \sqrt{ m_1^2 + m_2^2  + 2 (E_1 E_2 - \vec{p}_1 \cdot \vec{p}_2) } ,
  \label{one}
\end{equation}
where $m_{1,2}$, $\vec{p}_{1,2}$, $E_{1,2}$ are respectively the mass, momentum vector and energy of the decay products.
The derivative of this expression with respect to the magnitude of the momentum vector $p_1 = |\vec{p_1}|$ can be written as
\begin{equation}
  \frac{\deriv m_{12}}{\deriv p_1} \; = \; \frac{ m_{12}^2 - m_1^2 - m_2^2 - 2 m_1^2 E_2 / E_1}{ 2 \: m_{12} \: p_1 } .
\end{equation}
Consequently, if the momentum scale of child $d_1$ is biased by a scale factor
\begin{equation}
  p_1 \to (1 + \alpha_1) \: p_1.
\end{equation}
The observed change in the two-body invariant mass, to first order in $\alpha_1$, is given by
\begin{equation}
  m_{12} \to m_{12} + \alpha_1 \: \frac{ m_{12}^2 - m_1^2 - m_2^2 - 2 m_1^2 E_2 / E_1 }{2 m_{12}} .
  \label{equ:biassingledaughter}
\end{equation}
Therefore, the value of $\alpha_1$ can be obtained from the difference $\delta m$ between the observed average invariant
mass $m_{12}$ and the known mass of the parent particle $m_P$.

In general a momentum-scale bias will affect both child particles. To first order in the momentum-scale bias the total mass bias can be obtained by summing the two contributions. Assuming the momentum-scale bias is the same for both children (that is $\alpha_1=\alpha_2$), the observed bias in the mass is given by
\begin{equation}
  \delta m = \alpha \: \frac{ m_P^2 - m_1^2 - m_2^2 - m_1^2 E_2 / E_1 - m_2^2 E_1 / E_2}{m_P} .
\end{equation}
In the case of two identical child particles with mass $m$, the average bias is
\begin{equation}
  \delta m = \alpha \: \frac{ m_P^2 - m^2 R}{m_P},
  \label{eqnr}
\end{equation}
with
\begin{equation}
  R = 2 + \frac{E_1}{E_2} + \frac{E_2}{E_1} .
  \label{rform}
\end{equation}
Equation \ref{eqnr} is appropriate for decays such as $\KS\to\pi^+\pi^-$ where the mass of
the child particles is comparable to the mass of the parent. For the decays $\jpsi  \rightarrow \mu^+ \mu^- $ and $\PUpsilon(nS)
\rightarrow \mu^+ \mu^-$, measured in the LHCb detector, $m^2R \ll m_P^2$ and hence
\begin{equation}
\label{eq:jpsi}
\Delta m \approx \alpha \cdot m_P .
\end{equation}

The $\Bu \rightarrow \jpsi K^+$ decay mode\footnote{Charge
  conjugation is implied throughout.}, with the subsequent $\jpsi \rightarrow
\mu^+ \mu^-$ decay, is crucial for the calibration of the LHCb
spectrometer. For this decay, after applying a kinematic fit \cite{Hulsbergen:2005pu} that constrains the measured dimuon mass to the
known $\jpsi$ mass, the mass resolution is dominated by the
knowledge of the momentum of the kaon. Consequently, it is reasonable to assume that the reconstructed $\Bu$  mass is directly related to
the momentum scale of the charged kaon. Labelling the kaon momentum by $p_1$ the momentum-scale bias is obtained using the average value of the mass bias corrected by the kinematic factor in Eq.~\ref{equ:biassingledaughter},
\begin{equation}
  \alpha \; = \; (m_{12} - m_{P})  \: \frac{2 m_{12}}{ m_{12}^2 - m_1^2 - m_2^2 - 2 m_1^2 E_2 / E_1 } .
  \label{eqscale3}
\end{equation}
This method gives the momentum-scale in bins of the kaon kinematics. The momentum $p_2$ of the mass-constrained $\jpsi$ is considered to be unaffected by the momentum-scale bias.  Studies with the $\Bu \rightarrow \jpsi \Kp$ simulation indicate that this assumption is valid to a precision of $10^{-4}$ in the parameter $\alpha$.

A momentum-scale bias as discussed above is generated by a wrong estimation of the magnetic field integral or a poor tuning of the energy loss correction in the detector. Another possible effect is the curvature bias due to detector misalignment  discussed in Ref.~\cite{Amoraal:2012qn}. For the LHCb geometry, a typical misalignment that leads to a curvature bias is a small displacement, in the bending plane, of the T-stations relative to the VELO. Defining the curvature of particle $i$ as $\omega_i = q_i/p_i$, where $q_i$ is the particle charge, the effect of a curvature bias $\omega_1 \to \omega_1 + \delta\omega$ on the two-body invariant mass follows from Eq.~\ref{equ:biassingledaughter} as
\begin{equation}
\delta m_{12} = - \delta\omega \: \frac{p_1}{q_1} \: \frac{m_{12}^2 - m_1^2 - m_2^2 - 2m_1^2 E_2/E_1 }{2 m_{12}}.
\label{eq:curv}
\end{equation}

Contrary to a global momentum-scale bias, the mass bias due to a  curvature bias has opposite sign for
the particles in a two-body decay. If the two particles have equal momentum, the two contributions exactly cancel. Labelling the momentum of the negatively charged child as $p_1$, in the limit of massless children Eq.~\ref{eq:curv} becomes
\begin{equation}
 \delta m_{12} = \frac{1}{2} \delta\omega \: m_{12} \: (p_1 - p_2).
\end{equation}
Hence, a curvature bias can be identified by considering the average invariant mass as a function of the momentum difference between the children.

A curvature bias leads to a relative momentum bias that is quadratic in the momentum. As a result momentum-scale bias effects typically dominate at small momenta, while curvature bias effects dominate at large momenta. For the LHCb detector, the curvature bias from residual misalignment needs to be accounted for in electroweak physics measurements, such as the $W$-boson mass measurement detailed in Ref.~\cite{LHCb-PAPER-2021-024,LHCb-DP-2023-001}. Applying an additional correction for curvature bias effects has little impact on the results presented here.

\section{Radiative corrections and momentum scale calibration}
\label{sec:radiative}
\begin{figure}[h!]
\begin{center}
\includegraphics[width=0.48\textwidth]{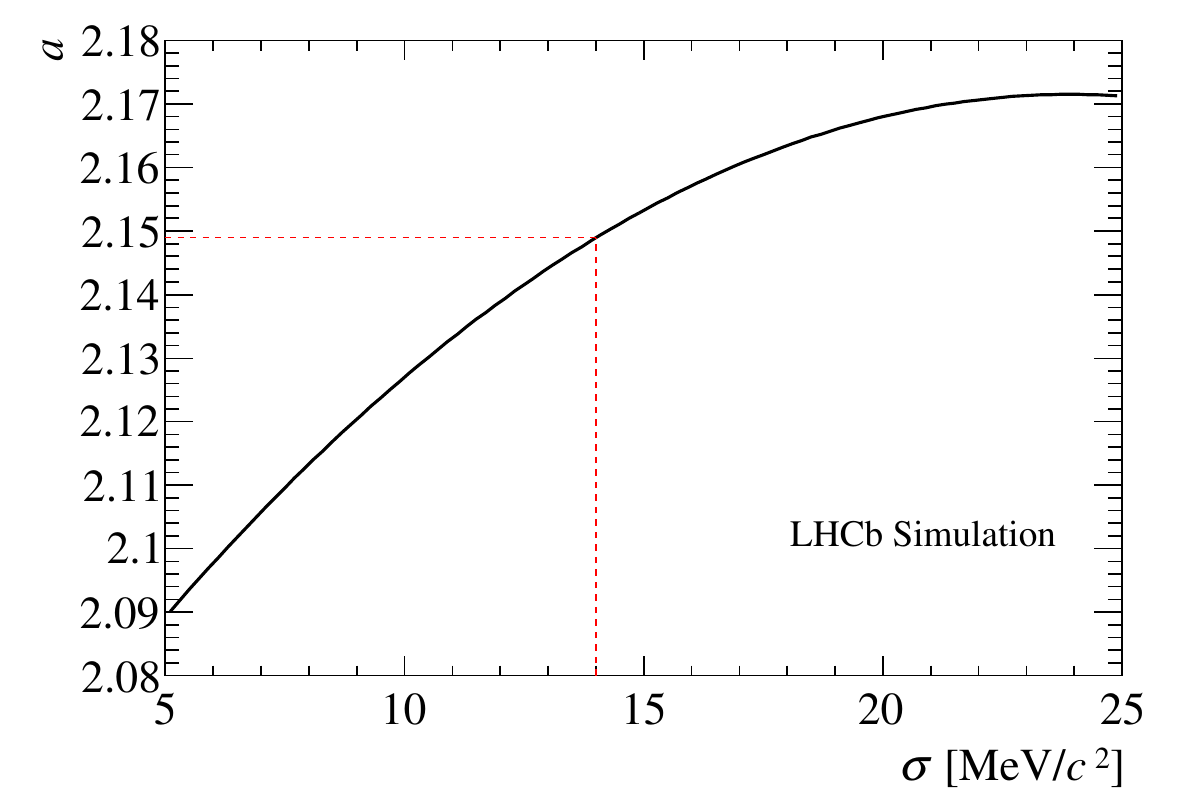}
\includegraphics[width=0.48\textwidth]{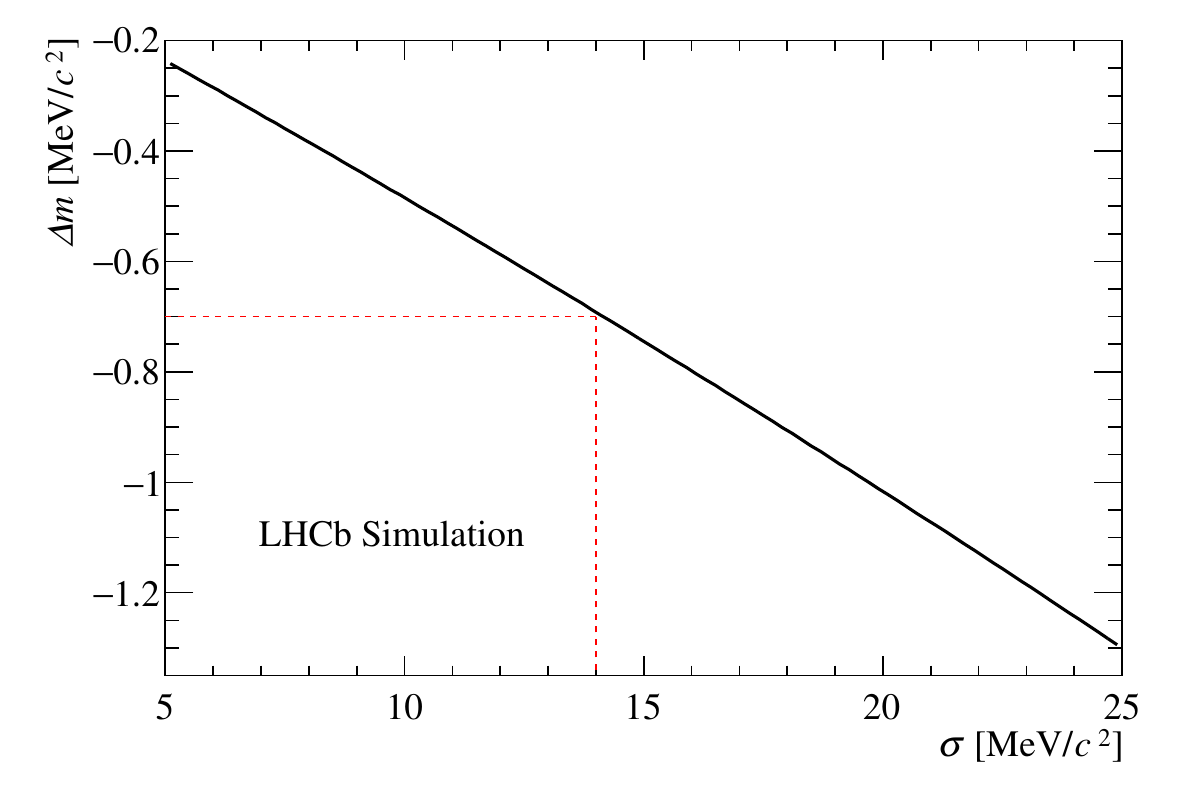}
\caption{\small (left) Parameterization of $a$ and (right) bias on the
  measured mass versus $\sigma$ for the $\jpsi \rightarrow
  \mu^+ \mu^-$ decay mode. The (red) dashed lines highlight the values for the
   typical $\jpsi$ mass resolution of $14 \mevcc$.
 }
\label{fig:rad}
\end{center}
\end{figure}

A complication that arises in the extraction of the momentum scale from
resonance decays is due to QED final state radiation. This has two
effects. First, it introduces a long radiative tail that must be
accounted for to obtain a good quality fit. A common approach is to
fit a Crystal Ball function \cite{Skwarnicki:1986xj} which has a
Gaussian core and a power-law tail. The
Crystal Ball function has four parameters: mean ($\mu$), resolution
($\sigma$), transition point for the radiative tail ($a$) and the
exponent of the power law ($n$). The latter three parameters are highly
correlated and care is needed to obtain a reliable fit. The procedure
adopted is to fix $n$ to one and parameterize $a$ from fits to the shape obtained from the generator-level simulation convolved with a Gaussian resolution function with known $\sigma$. As an example, the resulting quadratic function for the $\jpsi
\rightarrow \mu^+ \mu^-$ decay mode is shown in Fig. \ref{fig:rad}(left).

The second effect of final state radiation is to shift the position of
the mass peak to lower values. This bias remains even if a
fit to a Crystal Ball function is made since it has a symmetric Gaussian
core. It is corrected for by parameterizing
the observed bias in the simulation as a function of the
resolution. Figure~\ref{fig:rad}(right) shows the resulting correction for
the $\jpsi \rightarrow \mu^+ \mu^-$ decay mode. The typical correction size is $0.7 \mevcc$,
which corresponds to $\alpha \sim 2 \times 10^{-4}$. The reliability of the \photos simulation is probed by varying its settings leading to a relative uncertainty of $5 \times 10^{-5}$ on the momentum scale determination.

\section{Method}
\label{sec:method}
\begin{figure}[h!]
\begin{center}
\includegraphics[width=0.49\textwidth]{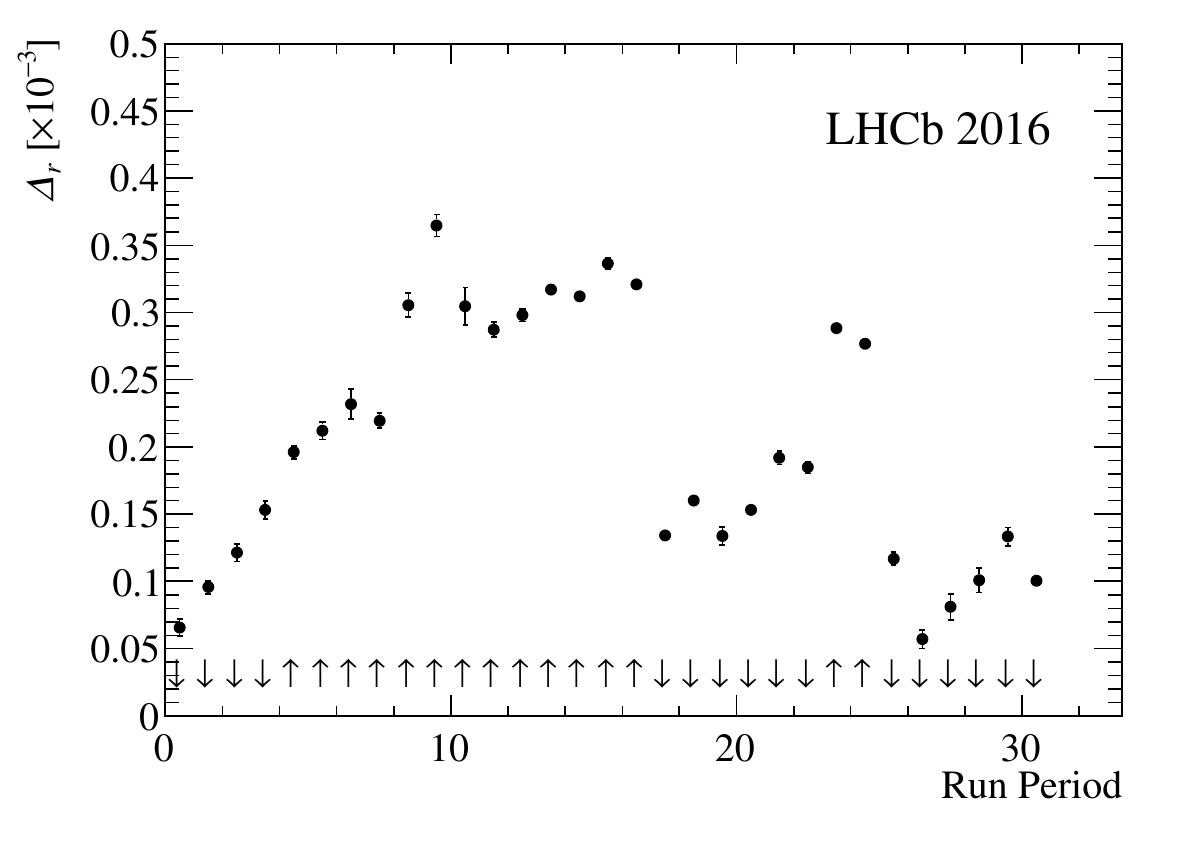}
\includegraphics[width=0.44\textwidth]{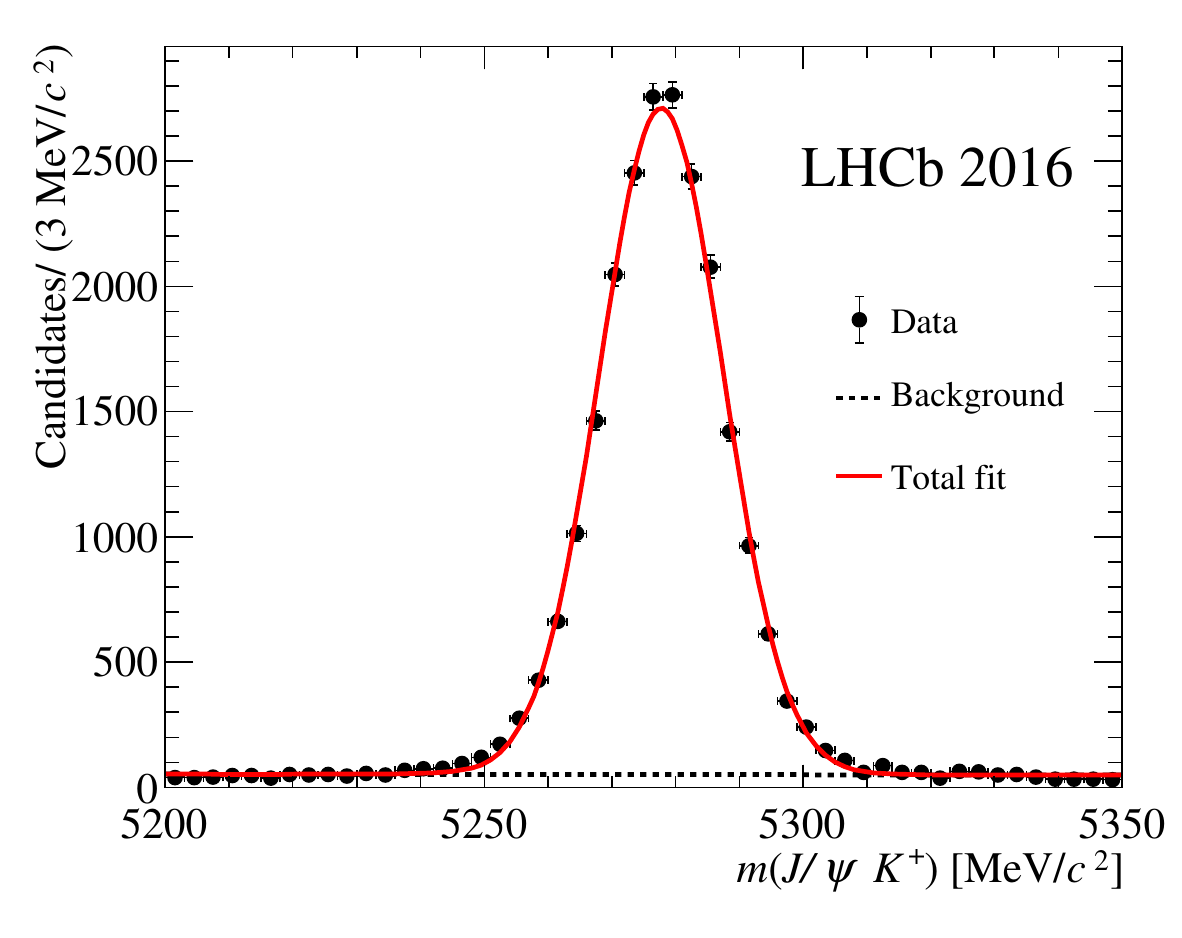}
\caption{\small (left) Variation of $\Delta_{r}$, during the 2016 run period determined using
  the $\jpsi \rightarrow \mu^+ \mu^-$ decay mode. The magnetic field polarity is indicated below the plot with the symbols $\uparrow,\downarrow$ for field up and down respectively. (right) Example fit to the $ J/\psi K^+$ invariant mass distribution for
  candidates with $q \cdot f_p > 0$ and $ 0.025 <
  tx < 0.05$~rad and $ 0 < ty < 0.05$~rad for the 2016 dataset.
 }
\label{fig:time}
\end{center}
\end{figure}

The procedure used to correct the momentum scale exploits the
large samples of detached $\jpsi \rightarrow \mu^+ \mu^-$ and
$\Bu \rightarrow \jpsi K^+$ decays collected by LHCb. The calibration is
performed separately for each year of data taking and is illustrated here
using data collected during 2016. Corrections are performed
that account for the variation of the momentum scale with time, $\Delta_r$,
particle kinematics, $\Delta_k$, and to fix the global momentum scale, $\Delta_g$. In the first step the data sample is divided into run
periods chosen according to known changes in conditions
(e.g. reversals of the magnet polarity) or known drifts in the measured
$\jpsi$ mass. For each run period a fit is made to the dimuon invariant mass
distribution of selected $\jpsi$ candidates. In this fit the signal is
modelled by a Crystal Ball function with the tail parameterisation
described in Section \ref{sec:radiative} and the background by an
exponential function. The fitted mean mass value is corrected for the effect of radiative corrections
and converted into $\Delta_r$ using Eq.~\ref{eq:jpsi}. In Fig. \ref{fig:time} (left) it can be seen that the size of $\Delta_r$ varies from $1-4 \times 10^{-4}$ over the 2016 running period. Two trends are visible. At the start of the 2016 run period a global upwards drift of $\Delta_r$ is observed. This is correlated with movements of the hardware monitoring systems of the downstream tracking stations over time \cite{LHCb-DP-2013-003}. In addition, $\Delta_r$ for data taken with \MagUp is generally larger than \MagDown by $\sim 10^{-4}$. A similar effect is observed in all LHCb running during Run 1 and 2. It may be due to the shifts in
the T-station positions when the field is reversed as discussed in
Section \ref{sec:Detector} or to movements of the magnet coils.

\begin{figure}[h!]
\begin{center}
\includegraphics[width=0.48\textwidth]{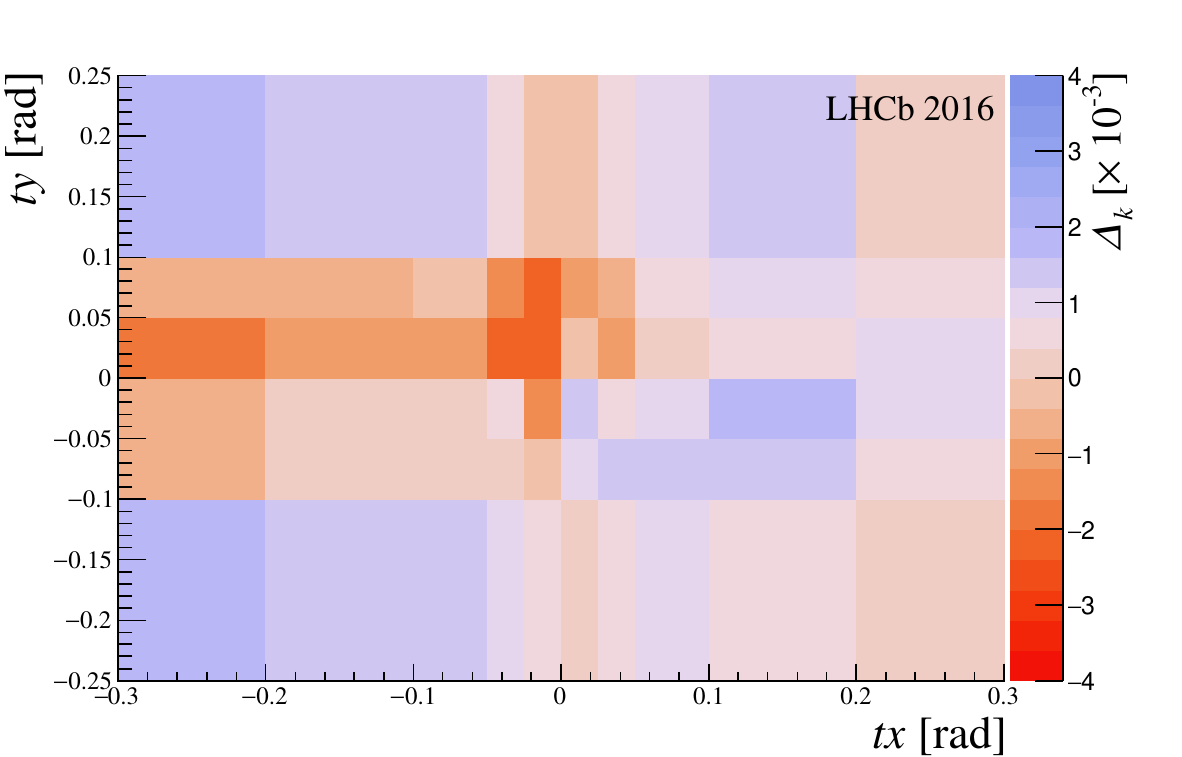}
\includegraphics[width=0.48\textwidth]{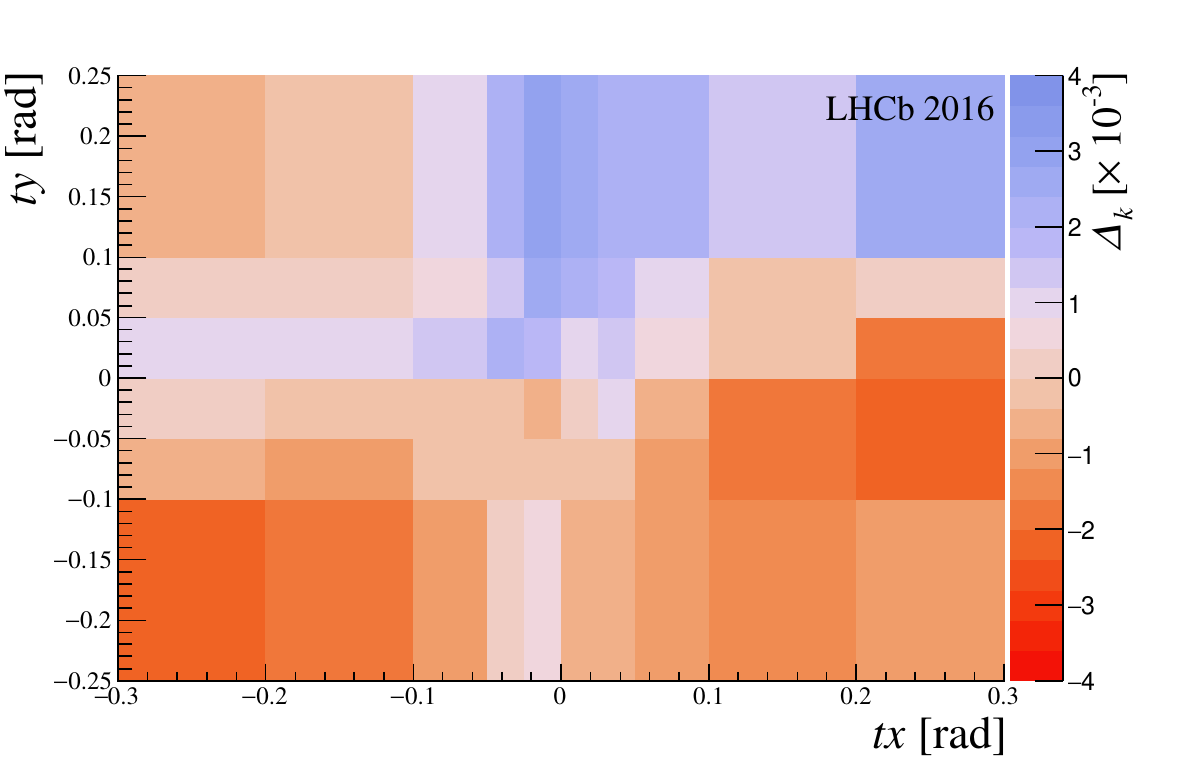}
\caption{\small Local momentum scale correction $\Delta_{k}$ as a function of  $tx$ and $ty$ for  the 2016 dataset (left) $q
  \cdot f_p > 0$ and (right) $q \cdot f_p < 0$.  }
\label{fig:cmaps}
\end{center}
\end{figure}

The next step is to determine $\Delta_k$. Fits are made to the invariant mass distribution of
$\Bu \rightarrow \jpsi K^+$ candidates in bins of the kaon kinematics
and the field scale determined by converting the fitted value of the  $\Bu$ mass
into an estimate of the momentum scale using Eq. \ref{eqscale3}.  Ideally, the phase space would be binned according to the kaon track
slopes ($tx = \textrm{tan}^{-1}(p_{x}/p_{z})$ and $ty = \textrm{tan}^{-1}(p_{y}/p_{z}$)) and curvature ($q/p$). With
the available dataset size this is not possible. Various binning
schemes and reduced sets of variables were explored. The best performance, judged by the improvement
in the mass resolution for the $\PUpsilon(nS)$ resonances, was obtained
by dividing the data according to the sign of the product of the track charge and magnetic
field polarity\footnote{$f_p = -1$ for \textit{MagDown} and 1 for \textit{MagUp}.}, $q \cdot f_p$, and binning according to $tx$ and $ty$. The binning scheme
for each year was adjusted such that statistical uncertainty on each
bin was $\mathcal{O}(10^{-4})$. An example mass fit to the $\Bu \rightarrow
\jpsi K^+$ mode is shown in Fig. \ref{fig:time} (right). The resulting
correction maps are shown in Fig. \ref{fig:cmaps}. Particularly at the
edge of the detector the values of $\Delta_k$ are large ($\Delta_k
\sim 10^{-3}$). Local corrections to the momentum scale at the level of $10^{-3}$ are needed.

\begin{figure}[htb!]
\begin{center}
\resizebox{3.5in}{!}{\includegraphics{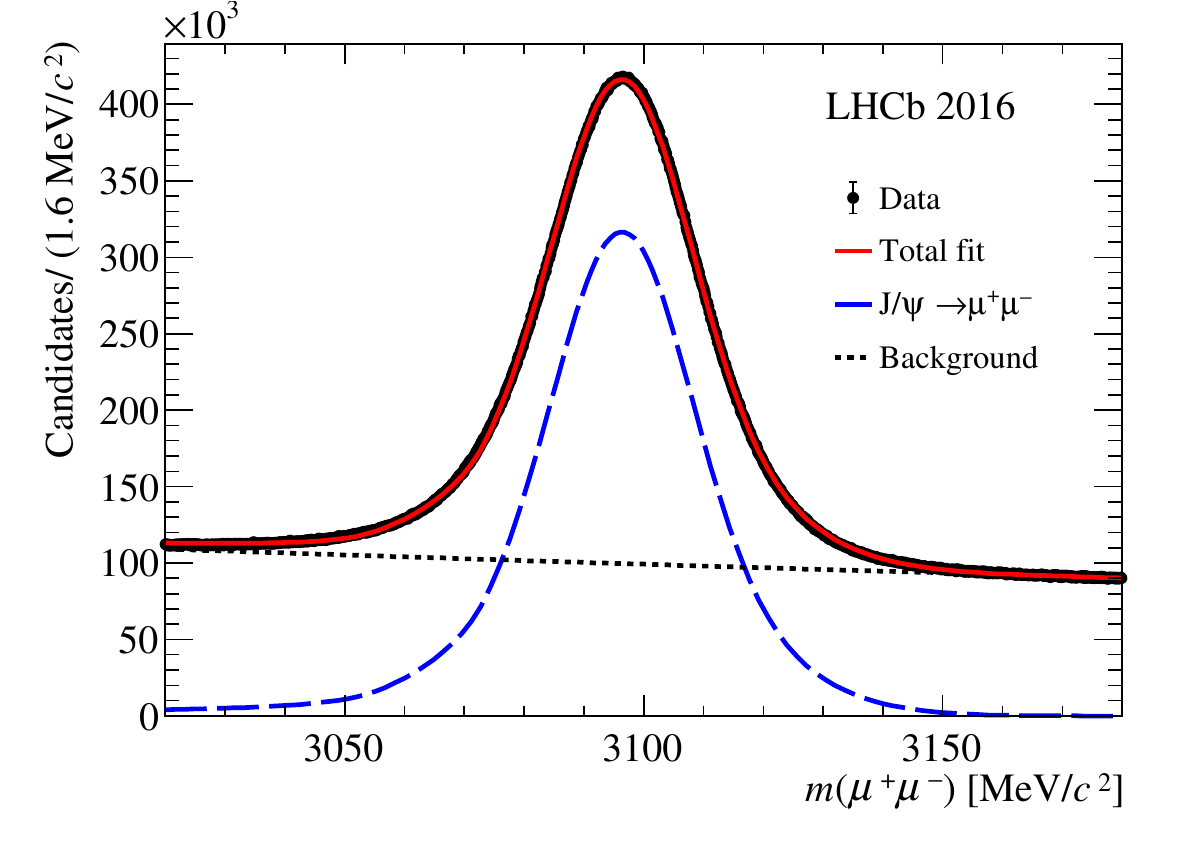}}
\caption{\small Fit to the dimuon invariant mass distribution for the
  $J/\psi$ sample after the procedure discussed in the text. The signal  (dashed blue line), combinatorial background (dotted black line) and total fit mode (red line) are superimposed. }
\label{fig4}
\end{center}
\end{figure}

The final step is to fix, $\Delta_g$, the global momentum scale. This is done by
fitting the  $\jpsi \rightarrow \mu^+ \mu^-$ dataset with a signal
described by the sum of
two Crystal Ball functions and an exponential background component. The
$J/\psi$ mass distribution after this calibration is shown in Fig.~\ref{fig4}.

The total momentum scale correction applied to each particle is thus
\begin{equation*}
1 - \alpha = 1 - \Delta_{g} - \Delta_{k}  - \Delta_{r}.
\end{equation*}

\section{Validation}
\label{sec:validation}
\begin{figure}[htb!]
\begin{center}
\includegraphics[width=0.49\textwidth]{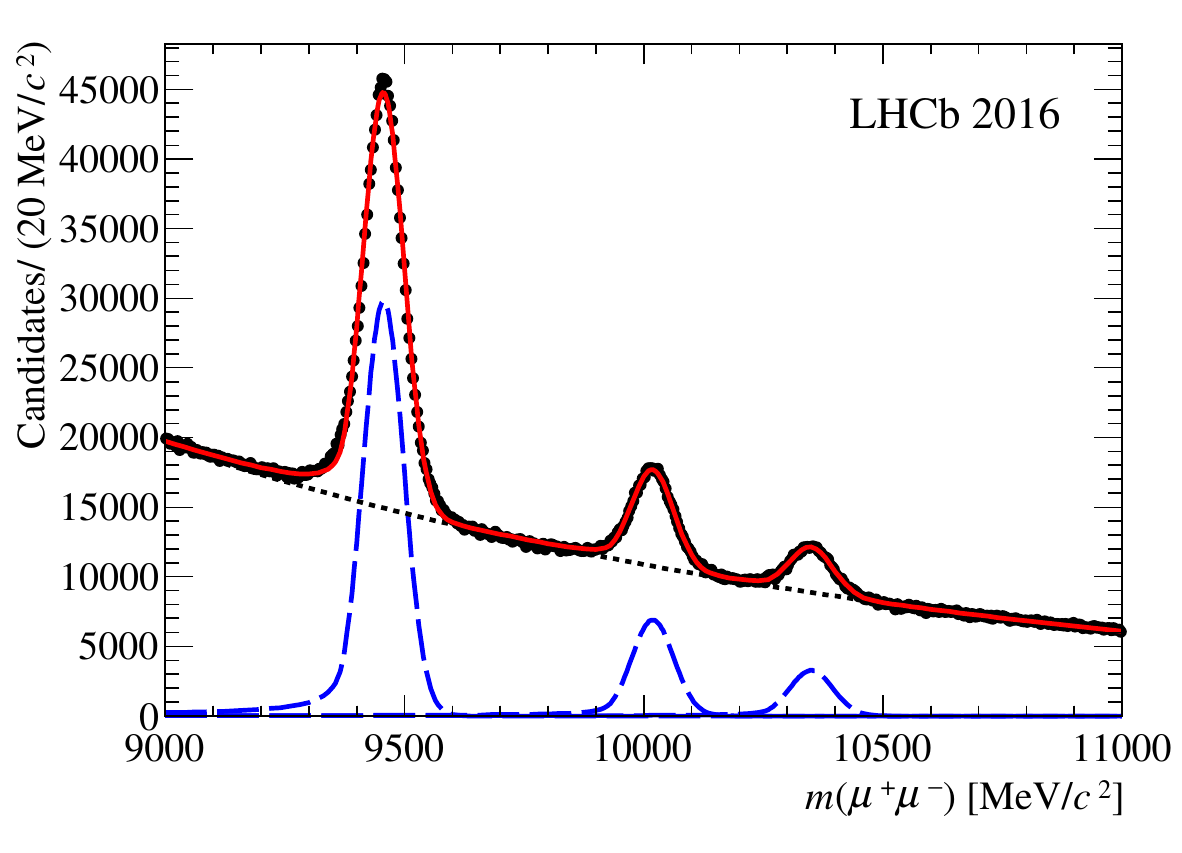}
\includegraphics[width=0.49\textwidth]{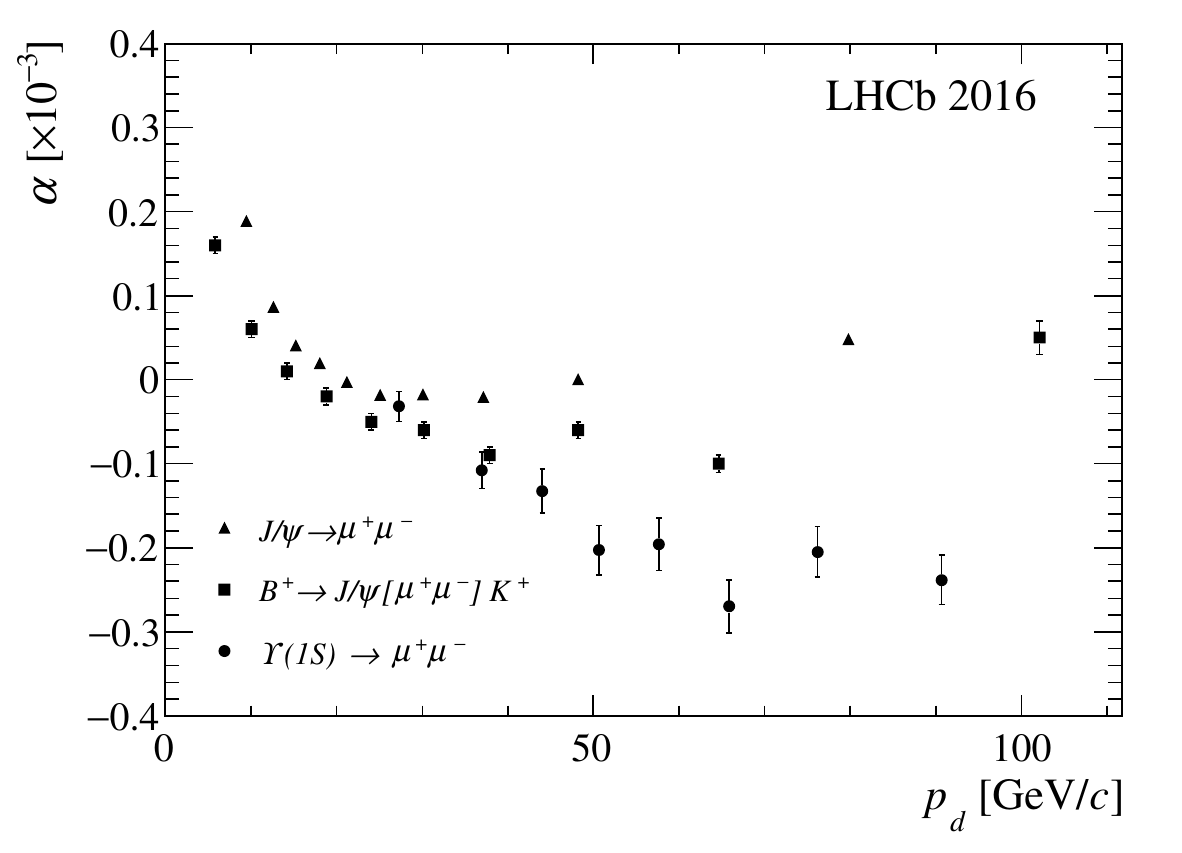}
\caption{\small  (left) Fit to the sample of selected dimuon candidates in the
  region of the $\PUpsilon$ resonances. Each  $\PUpsilon$ resonance is described by a Crystal Ball function (blue dashed line) and the combinatorial background by an exponential (black dotted line). The total fit model is also superimposed. (right) Bias on the momentum
  scale, $\alpha$ versus $p_d$ as defined in the text for the $\jpsi
  \rightarrow \mu^+ \mu^-$, $\Bu \rightarrow \jpsi K^+$ and the
  $\OneS \rightarrow \mu^+ \mu^-$ samples.}
\label{fig:val}
\end{center}
\end{figure}

Cross-checks are made of the validity of the calibration
procedure using the $\jpsi \rightarrow \mup \mun$ and
$\Bu \rightarrow \jpsi K^+$ samples together with a sample of $\PUpsilon(nS)$
decays shown in Fig. \ref{fig:val} (left).  The large samples available allow the variation of
the momentum scale to be studied as a function of kinematic variables such as $p$, transverse momentum and the orientation of the decay plane with respect to the magnetic field. As an example, Fig. \ref{fig:val} (right) shows the remaining bias on the momentum scale
as a function of $p_d$ which is the momentum of the kaon for the
$\Bu \rightarrow \jpsi K^+$ sample and half the momentum of the
parent in the case of the $\jpsi$ and $\OneS$ resonances. It can
be seen that the variation of $\alpha$ as a function of momentum is less than $3
\times 10^{-4}$.

The size of $\Delta_k$, is similar in magnitude to the
momentum resolution. As can be seen in Table \ref{tab:ures} applying
the calibration improves the relative mass resolution for the $\PUpsilon(nS)$
resonances by around $5 \%$.

\begin{table}[htb!]
\caption{\small Invariant mass resolution for the $\PUpsilon$
  resonances before and after the momentum scale calibration.  }
\begin{center}
\small
\begin{tabular}{c|l@{$\,\pm\,$}ll@{$\,\pm\,$}l}
& \multicolumn{4}{l}{Mass resolution [$\mevcc$]} \\
\raisebox{1.5ex}[-1.5ex]{Resonance} & \multicolumn{2}{c}{Before} &
                                                                   \multicolumn{2}{l}{\phantom{Ba}After}
  \\ \hline
$\OneS$ & 44.4 & 0.1 & 42.4 & 0.1\\
$\TwoS$ & 45.9 & 0.3  & 43.9 & 0.2 \\
$\ThreeS$ &47.3 & 0.5 &45.2 & 0.5\\

\end{tabular}
\end{center}
\label{tab:ures}
\end{table}

Various checks are made using the $\Bu \rightarrow \jpsi K^+$ sample. For
example, dividing the data according to whether the kaon has hits in
the TT detector, whether it traversed the inner or outer parts of the downstream
tracker and the location of its first measurements in the vertex
detector.  In these tests no evidence of a bias significantly larger
than the $3 \times 10^{-4}$ level was found in any kinematic region.

\begin{figure}[htb!]
\begin{center}
\resizebox{3.5in}{!}{\includegraphics{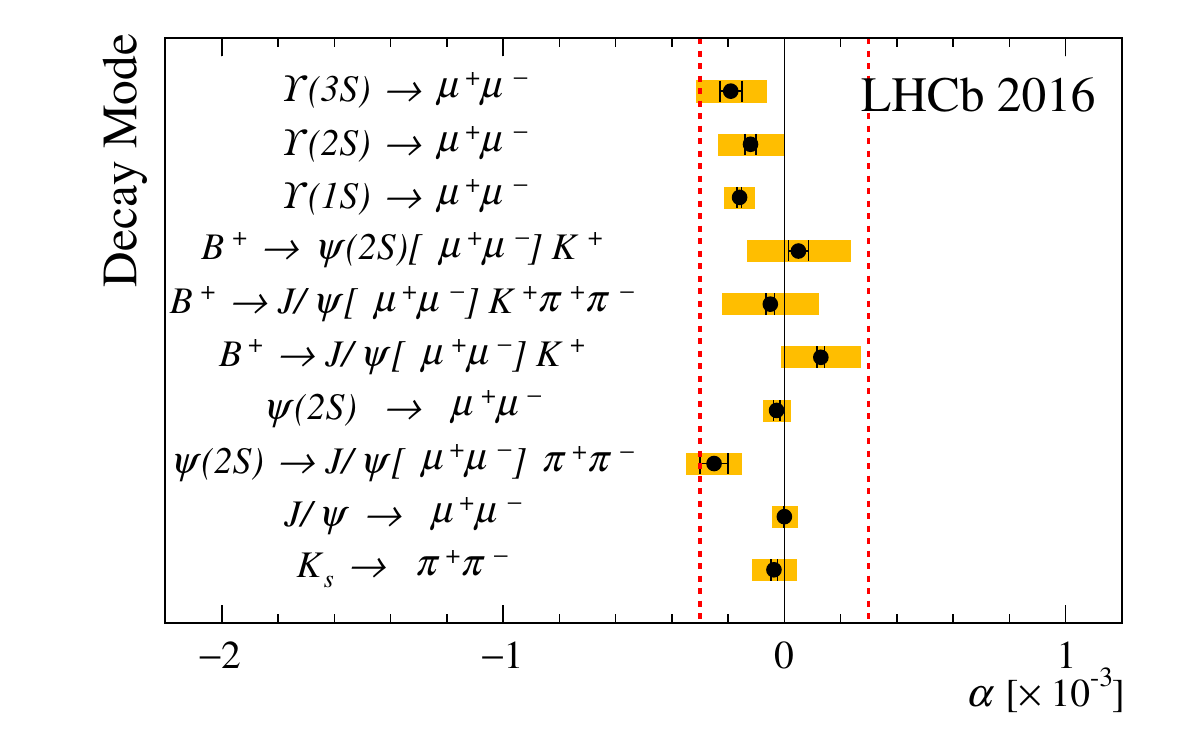}}
\caption{\small Average momentum-scale bias $\alpha$ determined from the
  reconstructed mass of various decay modes after the momentum
  calibration procedure for the 2016 dataset. The $\KS$ decays considered are those that occur in the
  vertex detector and have a flight distance of less than $5 \cm$. The
  black error bars represent the statistical uncertainty whilst the
  (yellow) filled areas also include contributions to the systematic
  uncertainty from the fitting procedure, the effect of QED radiative
  corrections, and the uncertainty on the mass of the decaying meson \cite{PDG2022}. The (red) dashed lines show the assigned uncertainty of $\pm\,3 \times 10^{-3}$ on the momentum scale.}
\label{fig6}
\end{center}
\end{figure}

Figure \ref{fig6} summarizes the residual bias observed for a variety of different decay
modes that probe a range of particle kinematics. It can be seen that
the variation in $\alpha$ between the different modes again does not exceed
$3 \times 10^{-4}$.

Based upon the observed variation of $\alpha$ seen for different decay-modes and
as function of particle kinematics the uncertainty on $\alpha$ is taken to be $3 \times 10^{-4}$. A similar conclusion is drawn by considering the variation in the observed $D^0$ mass in four-body decay modes \cite{LHCb-PAPER-2013-011}.

\section{Comparison with Run 1}
\label{sec:comp}
\begin{figure}[htb!]
\begin{center}
\resizebox{3.5in}{!}{\includegraphics{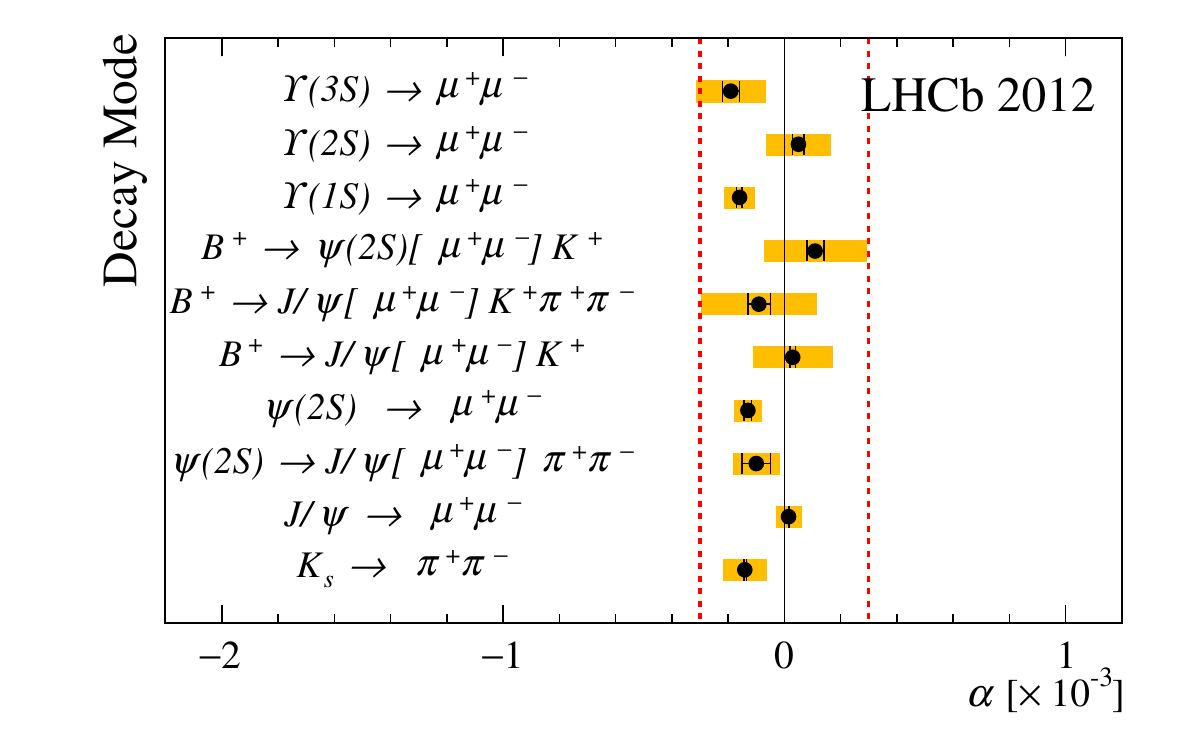}}
\caption{\small Average momentum-scale bias $\alpha$ determined from the
  reconstructed mass of various decay modes after the momentum
  calibration procedure for data collected in 2012. The $\KS$ decays considered are those that occur in the
  vertex detector and have a flight distance of less than $5 \cm$. The
  black error bars represent the statistical uncertainty whilst the
  (yellow) filled areas also include contributions to the systematic
  uncertainty from the fitting procedure, the effect of QED radiative
  corrections, and the uncertainty on the mass of the decaying meson. The (red) dashed lines show the assigned uncertainty of $\pm\,3 \times 10^{-3}$ on the momentum scale.}
\label{fig7}
\end{center}
\end{figure}

The procedure described here was used to calibrate Run 1 data (see for example
Ref. \cite{LHCb-PAPER-2013-011,LHCb-PAPER-2012-048}). The main difference between the two datasets is the size of the
overall correction. For the Run 1 data taking an earlier and less accurate version of the magnetic field map was
used.  With this field map values of $\Delta_r$ around $1 \times 10^{-3}$ were found. After the momentum scale calibration a similar performance to the 2016 data is
achieved. This can be seen in Figure \ref{fig7} where the residual
bias for a variety of decay modes is summarized for the study of the 2012 data.

\section{Selection biases}
\label{sec:bias}
An additional invariant mass bias is present for open charm decays selected with
cuts on quantities related to the flight distance. This can be seen by
considering a two-body decay such as $\Dz \rightarrow \Km
\pip$. If multiple scattering, that occurs within the RF-foil that separates the VELO from the primary LHC vacuum, increases the opening
angle between the decay products, both the reconstructed invariant mass
and the decay length will be higher than the true value. Thus if cuts are applied on the minimum flight
distance candidates with higher values of invariant mass are
selected. The size of this bias
depends on the selection and the lifetime of the decaying hadron. If
no lifetime biasing cuts are applied or the particles are prompt (such
as the $\PUpsilon(nS)$) or long-lived (like $\bquark$-hadrons), the
size of this effect is negligible. The lifetime of open charm hadrons
is such that with typical selections a large bias is seen. With the
selection applied in the trigger \cite{LHCb-PROC-2015-011} the size of
this bias is estimated using the simulation to be $0.7 \mevcc$  and
$0.2 \mevcc$ for the $\Dz$ and $\Dp$ mesons respectively. The
difference in the size of the bias reflects the difference in the
lifetime of the two mesons. Several approaches to correct this are adopted. For the studies of open
charm meson masses presented in Ref.~\cite{LHCb-PAPER-2013-011} a lifetime unbiased selection is used. In other analyses, such as
the discovery of the $\Xi_{cc}^{++}$ baryon \cite{LHCb-PAPER-2017-018}, tight requirements on lifetime related variables are needed. In such cases the bias is studied and corrected using the simulation.

\section{Summary}
\label{sec:summary}
In this paper the calibration of the momentum scale of the LHCb
detector has been described and illustrated using data collected
during 2016. The procedure used is generic and makes use of the large $\jpsi
\rightarrow \mup \mun$  and $\Bu \rightarrow \jpsi K^+$ samples that have been
collected. The uncertainty on the procedure is judged to be $3 \times 10^{-3}$ by studying the decays of well known resonances. Similar performance is achieved for the other years in Run 2. The method has allowed the LHCb collaboration to make many precise determinations of particle masses and widths over the last decade. Further improvements are possible with a better understanding of the field map and detector material.

\section*{Acknowledgements}
\noindent We express our gratitude to our colleagues in the CERN
accelerator departments for the excellent performance of the LHC. We
thank the technical and administrative staff at the LHCb
institutes.
We acknowledge support from CERN and from the national agencies:
CAPES, CNPq, FAPERJ and FINEP (Brazil);
MOST and NSFC (China);
CNRS/IN2P3 (France);
BMBF, DFG and MPG (Germany);
INFN (Italy);
NWO (Netherlands);
MNiSW and NCN (Poland);
MCID/IFA (Romania);
MICINN (Spain);
SNSF and SER (Switzerland);
NASU (Ukraine);
STFC (United Kingdom);
DOE NP and NSF (USA).
We acknowledge the computing resources that are provided by CERN, IN2P3
(France), KIT and DESY (Germany), INFN (Italy), SURF (Netherlands),
PIC (Spain), GridPP (United Kingdom),
CSCS (Switzerland), IFIN-HH (Romania), CBPF (Brazil),
and Polish WLCG (Poland).
We are indebted to the communities behind the multiple open-source
software packages on which we depend.
Individual groups or members have received support from
ARC and ARDC (Australia);
Key Research Program of Frontier Sciences of CAS, CAS PIFI, CAS CCEPP,
Fundamental Research Funds for the Central Universities,
and Sci. \& Tech. Program of Guangzhou (China);
Minciencias (Colombia);
EPLANET, Marie Sk\l{}odowska-Curie Actions, ERC and NextGenerationEU (European Union);
A*MIDEX, ANR, IPhU and Labex P2IO, and R\'{e}gion Auvergne-Rh\^{o}ne-Alpes (France);
AvH Foundation (Germany);
ICSC (Italy);
GVA, XuntaGal, GENCAT, Inditex, InTalent and Prog.~Atracci\'on Talento, CM (Spain);
SRC (Sweden);
the Leverhulme Trust, the Royal Society
 and UKRI (United Kingdom).

\section*{Acknowledgements}
\noindent We express our gratitude to our colleagues in the CERN
accelerator departments for the excellent performance of the LHC. We
thank the technical and administrative staff at the LHCb
institutes.
We acknowledge support from CERN and from the national agencies:
CAPES, CNPq, FAPERJ and FINEP (Brazil);
MOST and NSFC (China);
CNRS/IN2P3 (France);
BMBF, DFG and MPG (Germany);
INFN (Italy);
NWO (Netherlands);
MNiSW and NCN (Poland);
MCID/IFA (Romania);
MICINN (Spain);
SNSF and SER (Switzerland);
NASU (Ukraine);
STFC (United Kingdom);
DOE NP and NSF (USA).
We acknowledge the computing resources that are provided by CERN, IN2P3
(France), KIT and DESY (Germany), INFN (Italy), SURF (Netherlands),
PIC (Spain), GridPP (United Kingdom),
CSCS (Switzerland), IFIN-HH (Romania), CBPF (Brazil),
and Polish WLCG (Poland).
We are indebted to the communities behind the multiple open-source
software packages on which we depend.
Individual groups or members have received support from
ARC and ARDC (Australia);
Key Research Program of Frontier Sciences of CAS, CAS PIFI, CAS CCEPP,
Fundamental Research Funds for the Central Universities,
and Sci. \& Tech. Program of Guangzhou (China);
Minciencias (Colombia);
EPLANET, Marie Sk\l{}odowska-Curie Actions, ERC and NextGenerationEU (European Union);
A*MIDEX, ANR, IPhU and Labex P2IO, and R\'{e}gion Auvergne-Rh\^{o}ne-Alpes (France);
AvH Foundation (Germany);
ICSC (Italy);
GVA, XuntaGal, GENCAT, Inditex, InTalent and Prog.~Atracci\'on Talento, CM (Spain);
SRC (Sweden);
the Leverhulme Trust, the Royal Society
 and UKRI (United Kingdom).

\addcontentsline{toc}{section}{References}
\ifx\mcitethebibliography\mciteundefinedmacro
\PackageError{LHCb.bst}{mciteplus.sty has not been loaded}
{This bibstyle requires the use of the mciteplus package.}\fi
\providecommand{\href}[2]{#2}

\newpage

\centerline
{\large\bf LHCb collaboration}
\begin
{flushleft}
\small
R.~Aaij$^{35}$\lhcborcid{0000-0003-0533-1952},
A.S.W.~Abdelmotteleb$^{54}$\lhcborcid{0000-0001-7905-0542},
C.~Abellan~Beteta$^{48}$,
F.~Abudin{\'e}n$^{54}$\lhcborcid{0000-0002-6737-3528},
T.~Ackernley$^{58}$\lhcborcid{0000-0002-5951-3498},
B.~Adeva$^{44}$\lhcborcid{0000-0001-9756-3712},
M.~Adinolfi$^{52}$\lhcborcid{0000-0002-1326-1264},
P.~Adlarson$^{78}$\lhcborcid{0000-0001-6280-3851},
C.~Agapopoulou$^{46}$\lhcborcid{0000-0002-2368-0147},
C.A.~Aidala$^{79}$\lhcborcid{0000-0001-9540-4988}
% ...

{\footnotesize \it

% ...
$^{35}$Nikhef National Institute for Subatomic Physics, Amsterdam, Netherlands\\
% ...
$^{44}$Instituto Galego de F{\'\i}sica de Altas Enerx{\'\i}as (IGFAE), Universidade de Santiago de Compostela, Santiago de Compostela, Spain\\
% ...
$^{46}$European Organization for Nuclear Research (CERN), Geneva, Switzerland\\
% ...
$^{48}$Physik-Institut, Universit{\"a}t Z{\"u}rich, Z{\"u}rich, Switzerland\\
% ...
$^{52}$H.H. Wills Physics Laboratory, University of Bristol, Bristol, United Kingdom\\
% ...
$^{54}$Department of Physics, University of Warwick, Coventry, United Kingdom\\
% ...
$^{57}$School of Physics and Astronomy, University of Glasgow, Glasgow, United Kingdom\\
$^{58}$Oliver Lodge Laboratory, University of Liverpool, Liverpool, United Kingdom\\
% ...
$^{66}$Syracuse University, Syracuse, NY, United States\\
% ...
$^{78}$Department of Physics and Astronomy, Uppsala University, Uppsala, Sweden, associated to $^{57}$\\
$^{79}$University of Michigan, Ann Arbor, MI, United States, associated to $^{66}$\\
% ...
}
\end{flushleft}

\end{document}